\documentstyle[epsf]{article}
\newcommand{\EQ}{\begin{equation}}
\newcommand{\EN}{\end{equation}}

\newcommand{\authone}[2]{#1,#2}
\newcommand{\authtwo}[4]{#1,#2, and ~#3,#4}
\newcommand{\auththr}[6]{#1,#2,#3,#4, and ~#5,#6}
\newcommand{\authfour}[8]{#1,#2,#3,#4,#5,#6, and ~#7,#8} 
\newcommand{\authmanytwo}[4]{#1,#2,#3,#4,} 
\newcommand{\authmanythr}[6]{#1,#2,#3,#4,#5,#6,} 
\newcommand{\authmanyfour}[8]{#1,#2,#3,#4,#5,#6,#7,#8,} 

\newcommand{\yepl}[5]{ ``#5.'' Europhys. Lett.  {\bf #2}, #3-#4 (#1).}

\newcommand{\yjfm}[5]{ ``#5.'' J. Fluid Mech.  {\bf #2}, #3-#4 (#1).}

\newcommand{\yprl}[5]{ ``#5.'' Phys. Rev. Lett.  {\bf #2}, #3-#4 (#1).}
\newcommand{\ypra}[5]{ ``#5.'' Phys. Rev. A.  {\bf #2}, #3-#4 (#1).}
\newcommand{\ypre}[5]{ ``#5.'' Phys. Rev. E.  {\bf #2}, #3-#4 (#1).}

\newcommand{\yprsa}[5]{ #5  Proc. Roy. Soc. Lond. A
{\bf #2}, #3-#4 (#1).}

\newcommand{\ypf}[5]{ ``#5.'' Phys. Fluids  {\bf #2}, #3-#4 (#1).}
\newcommand{\ypfa}[5]{ ``#5.'' Phys. Fluids A {\bf #2}, #3-#4 (#1).}

\newcommand{\yblm}[5]{ ``#5.'' Boun. Lay. Met.  {\bf #2}, #3-#4 (#1).}
\newcommand{\yjour}[6]{ ``#6'' #2 {\bf #3}, #4-#5 (#1).}

\newcommand{\ybook}[3]{ {\em #2} (#3, #1).}
\epsfverbosetrue
\begin{document}

\title{An inertial range length scale in structure functions.}
\author{Robert M. Kerr\\
Geophysical Turbulence Program\\
National Center for Atmospheric Research\\
Boulder, CO 80307-3000, USA\\
Maurice Meneguzzi\\
ASCI-CNRS, Universite de Paris Sud, \\
91405 ORSAY Cedex and \\
Service d'Astrophysique, \\
Centre d'Etudes de Saclay, France\\
Toshiyuki Gotoh\\
Department of Systems Engineering, Nagoya Institute of Technology \\
Showa-ku, Gokiso-cho, Nagoya 466-8555, Japan}

\maketitle

\begin{abstract}
It is shown using experimental and numerical data that 
within the traditional inertial subrange
defined by where the third order structure function is linear that
the higher order structure function scaling exponents 
for longitudinal and transverse structure functions converge
only over larger scales, $r>r_S$, where $r_S$ has scaling
intermediate between $\eta$ and $\lambda$ as a function of $R_\lambda$.
Below these scales, scaling exponents cannot be determined for
any of the structure functions without resorting to procedures
such as extended self-similarity (ESS).  With ESS, different
longitudinal and transverse higher order 
exponents are obtained that are consistent with earlier results.
The relationship of these statistics to derivative and pressure
statistics, to turbulent structures and to length scales is discussed.
\end{abstract}

\section{Introduction \label{sec:intro}}
An important tool in understanding intermittency in turbulence
has been the exponents $\zeta_p$ of power laws 
for the velocity structure functions.
The longitudinal structure functions are
\EQ
S^L_p(r)=<(u(x+r)-u(x))^p>\sim r^{\zeta^L_p}
\label{eq:SL}
\EN
where $\overrightarrow{u}$ and $\overrightarrow{r}$ are in the same direction.  
Their measurement
requires only a single hot wire probe that can, through the Taylor
frozen turbulence assumption, determine one velocity component $u$
as a function of the parallel spatial direction $r$ and thus find the
$S^L_p$ at high Reynolds numbers.
Using crossed-wire probes, one can also obtain
reliable, high Reynolds number measurements of 
transverse structure functions $S^T_p(r)$,
where $\overrightarrow{u}$ and $\overrightarrow{r}$ are orthogonal,
and their exponents $\zeta^T_p$. 
There are also mixed structure functions containing both longitudinal and 
transverse components.  

The relationship between the $\zeta_p$ and intermittency \cite{K62}
is in deviations of
the the exponents $\zeta_p$ from their Gaussian or classical values
of $\zeta_p=p/3$, where $\zeta_2\approx2/3$ and $\zeta_3=1$ are
expected for an energy cascade.  
In the presence of intermittency, $\zeta_p<p/3$ for $p>3$ is expected.
Furthermore, it has generally been believed that all of the 
$\zeta^{L,T}_p$ of a given order $p$ should be 
the same in the infinite Reynolds number limit.  This is
closely related to the refined similarity hypothesis (RSH)
which assumes that the only 
information that can affect the statistics on a given scale $r$ is the
fluctuations in the energy cascade $\epsilon_r$ through that scale.
Details about either the
large scale forcing or the dissipative structures should be irrelevant
in this picture.  Measurements \cite{StolovKS92} have confirmed RSH 
as it relates dissipation to longitudinal statistics.

More recently, moderate Reynolds numbers simulations and experiments
\cite{BoratavP97,vandeWater95} have found, with the help of the extended
self-similarity hypothesis \cite{BenziESS}, that 
$\zeta^T_p~<~\zeta^L_p$ for $p>3$.  This has been now been
confirmed over all Reynolds numbers simulated \cite{Chen_SNC97} 
or observed \cite{DhruvaTsujiSreeni97} to date.
If longitudinal and transverse statistics
are different, then the statistics  of strain, that is dissipation 
$\epsilon_r$, and the statistics of 
of vorticity, call it $\Omega_r$ for its relationship
to the enstrophy, should be different.  Since $\zeta^T_p~<~\zeta^L_p$,
it implies that vorticity is more intermittent than strain.
This possibility has been
suggested by recent numerical results \cite{Chen_SNC97} that show that
dissipation and enstrophy statistics scale as RSH predicts, but
with separate distributions.  This result implies that 
when the nonlinear coupling is written as a convolution
in Fourier space that not only the velocity magnitude, but
also the velocity phase, what makes the transfer of 
$\Omega_r$ different than $\epsilon_r$,
is important.  Related to this, if the dissipation-dissipation correlation
and its $\Omega_r$ counterpart go as
\EQ
<\epsilon_x\epsilon_{x+r}>~\sim~r^\mu\quad{\rm and}
<\Omega_x\Omega_{x+r}>~\sim~r^{\mu^\Omega}
\label{eq:ddiss}
\EN
then different $\Omega_r$ statistics implies that $\mu^\Omega\neq\mu$.
Experimentally, $\mu\approx0.25$ \cite{Nelkin81}
has been confirmed by the latest measurements and simulations 
\cite{SreeniKai93,Pras_sp,Brachet91}, although lower Reynolds number
measurements tend to give $\mu=0.5$.
$\mu^\Omega\neq\mu$ is found in one numerical result \cite{Brachet91}.

It will be shown here that different longitudinal and transverse
statistics suggest the existence of a crossover length scale within the
inertial subrange.  That is a length scale distinct from 
the large length scale $L$ or a multiple of the dissipation or
Kolmogorov scale $\eta$, and maybe the order of the intermediate 
Taylor microscale $\lambda$, where the Taylor microscale Reynolds number 
and $\lambda$ are defined as
\EQ
R_\lambda={{U\lambda}\over\nu}\quad\quad{\rm where}\quad 
\lambda={U\over{<(du/dx)^2.>^{1/2}}}\label{eq:Rlambda}\EN
$R_\lambda$ is the definition of the Reynolds number that appears to give
uniform scaling in a variety of different flows and
is related to the large scale Reynolds number by $R_\lambda\sim(R=UL/\nu)^{1/2}$.
The Kolmogorov scale is related to $\lambda$ by
\EQ
\eta=(\nu/\epsilon)^{1/4}\sim \lambda R_\lambda^{-1/2}
\label{eq:eta}\EN
where $\nu$ is viscosity and $\epsilon$ is the dissipation rate.

In order to discern if there is some crossover length scale
for some structure function within the inertial subrange 
and determine if there are indeed different longitudinal
and transverse exponents, a long inertial subrange
and clean data are needed.  Numerical simulations do not have this
range, but if run sufficiently long give clean data
and more flexibility.  Observations can provide more
dynamic range, but with more limited types of data.
This paper will use both experimental and numerical data to try
to present a more complete picture than either measurements or
simulations alone of what evidence there is for
different longitudinal and transverse statistics and its implications. 
Experimental data will be used to indicate the high Reynolds number
trends, then analysis from forced numerical turbulence
in a box at $R_\lambda=262$ on a $512^3$ mesh and $R_\lambda=390$
on a $1024^3$ mesh will be presented, with most of the discussion related to the
$R_\lambda=262$ calculation.  It will be shown that if the same inertial
range scaling analysis defined by the experiments is applied to the simulations,
then the trends in the lower Reynolds number simulations are consistent
with the experiments and clearly demonstrate a trend where
simple scaling breaks down at many multiples of the Kolmogorov scale
within the traditional inertial subrange, and that this scale
appears to increase with Reynolds number.  

This paper will be organized as follows.  First, constraints on and
relationships between structure functions will be discussed.
Recent results on structure function and pressure structure
function scaling will be discussed in the context that different
scalings for the same order imply 
the existence of a statistically signficant
length scale within the inertial subrange where scaling properties
of at least some structure functions could change.  Next, there
will be a discussion of existing experimental and
observational results before showing new analysis of experimental data
up to $R_\lambda=3200$.  The analysis will use 
$S_2^{L,T}$ and $S_3$ to
define the minimum limits over which inertial subrange scaling analysis
can be applied before the results for the $\zeta^{L,T}_p$ for $p=4$
and 6 are presented. Then the numerical analysis will be presented
in a similar manner.


\section{Constraints \label{sec:constraints}}

\begin{figure}[htbp]
\vspace{-10pt}
\epsfbox{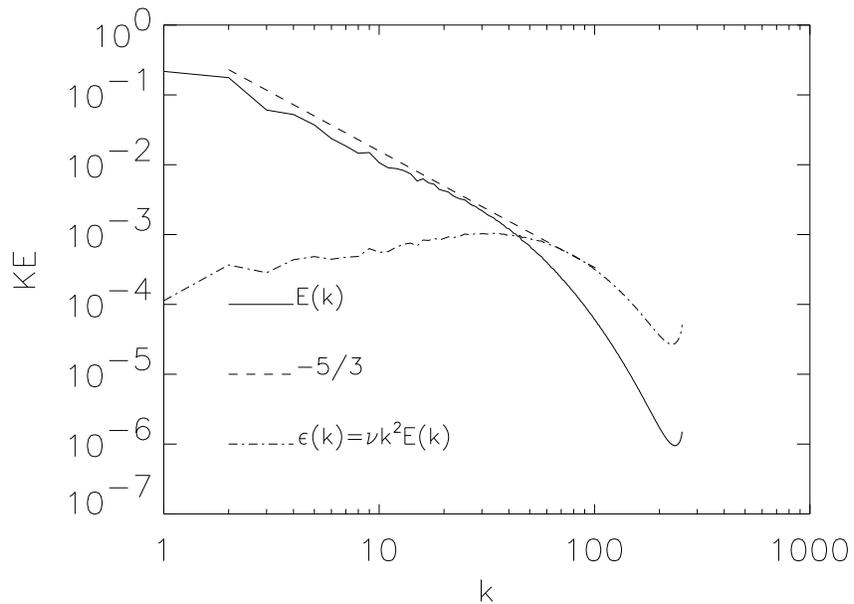}
\caption[]{Three-dimensional kinetic energy spectrum
$E(|k|)$ for the $512^3$ forced calculation.  The dashed line has 
an exponent of -5/3 and $k_\eta=k_{\max}=256$.  The dissipation
spectrum $\epsilon(k)=\nu k^2E(k)$ with a peak at $k=34$ is shown.
Resolution is determined
by the ratio of the wavenumber of the peak of the dissipation spectrum
(here $k\approx k_\eta/30=34$) to $k_{\max}$.  Experience has shown that
a ratio of at least 5 provides adequate resolution.}
\label{fig:sp512}
\end{figure}

This section discusses kinematical constraints upon structure functions,
related quantities, and what can be measured.  
There are only a few hard constraints for the scaling of 
small $p$ structure functions.  For $p=2$, assuming isotropy
and homogeniety and for a long enough
inertial subrange, $\zeta^L_2=\zeta^T_2$ is expected.
The classical value for $\zeta_2$ is 2/3 (\ref{eq:K41}), which
is closely related to assuming that the energy spectrum goes as
\EQ
E(k) =c_k\epsilon^{2/3}k^{-5/3}
\label{eq:K41}\EN
where $c_K$ is the experimentally determined Kolmogorov constant and
$\epsilon$ is the equivalently either the rate of energy dissipation
or the rate of energy transfer to small scales. With crossed wires,
only one-dimensional longitudinal or transverse spectra can be found, 
but with a numerical simulation the full energy spectrum can be obtained in
three-dimensions.  An example from the $512^3$ forced data
set to be discussed is shown in Figure \ref{fig:sp512}.
This spectrum is found
between a lower wavenumber, large-scale, forcing or integral scale $L$
and a high wavenumber, small-scale, dissipation scale 
$\eta=(\nu/\epsilon)^{1/4}$,
with $k_\eta=1/\eta$.  

\epsfxsize=12cm
The theoretical basis for predicting a -5/3 spectrum 
is the presumption of a local, uniform energy cascade.  
There is no constraint requiring -5/3 or the corresponding $\zeta_2=2/3$
for structure functions and corrections to the $\zeta_2=2/3$ are claimed
\cite{Anselmetetal84}.  
However, no corrections to a -5/3 energy spectrum have been
seen in numerical spectrum such as Figure \ref{fig:sp512}
and for the largest range of scales observationally \cite{Pras_sp}.
How could two supposedly equivalent measures of turbulent scaling,
the energy spectrum and the 2nd order structure functions, yield
two inconsistent results concerning intermittency?  A suggestion
below is that dissipation range effects on $S^L_2$ are more persistent than
thought.

For $p=3$, by balancing transfer and 
dissipation terms in the Karman-Howarth equation
\cite{K41,Frisch95}, $\zeta_3\equiv 1$ 
is expected. 
This is the only fixed constraint on the $\zeta_p$ for isotropic, homogeneous
turbulence regardless of intermittency and
yields Kolmogorov's 4/5 law for $S_3(r)$
\EQ
S_3(r)=-{4\over5}\epsilon r
\label{eq:K45th}\EN
However, linear in
$r$ behavior is never observed exactly (see discussion
with figure \ref{fig:3rd}).  This effect has been quantified
\cite{MTabelingW99} and it was shown that the peak of 
$-S_3/r$, after being compensated for the effects of forcing
and dissipation, has a Reynolds number dependence 
like $l_s/L \sim R_\lambda^{-3/5}$, whereas 
$\lambda/L\sim R_\lambda^{-1}$.  This could
suggest the existence of a dynamically significant position in
the inertial subrange that does not scale with either the small
Kolmogorov scale or the large scale and has been interpreted in terms of
an enstrophy production argument \cite{Novikov93}.

While it is difficult to obtain a clean power law for 
$p=3$, it is even harder to get convincing
power laws when plotting $S_p$ versus $r$ for $p>3$.
A device that has been used to determine higher order $\zeta_p$
is to make the assumption that even
if $-S_3\sim r$ is not exact, if it is assumed to be exact then
much stronger power laws for the $S_p$, $p\geq4$, can be obtained by
plotting $S_p$ versus $-S_3$ rather than versus $r$.
This is known as extended self-similarity or ESS \cite{BenziESS}.
Another way of looking at ESS is that the $S_p$ are not true
power laws all the way to the dissipation scale $\eta$, but
only for $r>A\eta$, where $A$ is large, and ESS allows one to
extend scaling much closer to $\eta$.  However, there is a question
of how much of this is a mirage because the ESS scaling exponents
as $r\rightarrow0$ are trivially the classical values of $p/3$.  
The additional length
scale $A\eta$ probably should have no dynamical significance since it is a
multiple of $\eta$, as will be discussed in section \ref{sec:why}.

For $p\geq 4$, there are no rigorous constraints that
the longitudinal and transverse structure functions should have the
same $\zeta_p$. However, it has been shown \cite{Siggia81a} 
that when all components of
the velocity field $\overrightarrow{u}$ and all directions of
$\overrightarrow{\partial_x}$ are considered that there are
four rotationally invariant combinations of 
fourth-order derivative correlations that can be 
written entirely in terms of the strain $e_{ij}$ and the
vorticity $\omega_i$ and have been discussed in detail \cite{Kerr85}. 
\EQ
F_e=(15/7){{<(e^2)^2>}\over{<e^2>^2}}\quad\quad
F_{e\omega}=3{{<\omega^2 e^2>}\over{<\omega^2><e^2>}}
\label{eq:Fe4}
\EN
$$
F_{\omega ee\omega}=3{{<\omega_i
e_{ij}e_{jk}\omega_k>}\over{<\omega^2><e^2>}}\quad\quad
F_\omega=(9/5){{<\omega^4>}\over{<\omega^2>^2}} $$
Under the assumptions of homogeneity and isotropy $F_e$ is equivalent
to the longitudinal fourth-order derivative flatness $F_4^L$ that can be
measured with a single hot-wire probe and is related to
the $r\rightarrow0$ limit of $S_4^L(r)$.  
To determine the other irrotational
flatnesses as functions of $R_\lambda$ can only be done at low Reynolds
numbers with complicated probes \cite{Tsinober} or 
simulations \cite{Kerr85,VincentM91},
which suggest that each has its own scaling with Reynolds number.

With a crossed-wire probe, higher Reynolds number observations of
the scaling of flatnesses related to
the $r\rightarrow0$ limit of the transverse and mixed structure functions
could be determined.  In addition to 
\EQ
F^L_4=F_{\ell\ell\ell\ell}=<u_{1,1}^4>/<u_{1,1}^2>^2=F_e
\label{eq:Fu4}
\EN
these are
\EQ
F^M_4=F_{\ell\ell tt}={{<u_{1,1}^2u_{2,1}^2>}\over{<u_{1,1}^2><u_{2,1}^2>}}
\label{eq:Fuv}
\EN
and
\EQ
F^T_4=F_{tttt}={{<u_{2,1}^4>}\over{<u_{2,1}^2>^2}}
\label{eq:Fv4} \EN
which can be related to combinations of the irrotational flatnesses (\ref{eq:Fe4}).
The scaling of $F^M_4$ and $F^T_4$ should be dominated
by their most intermittent components, which the analysis of
the irrotational components indicates is $F_\omega$ \cite{Kerr85}.

Generalized structure functions
should involve not only more than one velocity direction, but also
more than one spatial position and direction. 
All of the structure functions discussed so far involve different
velocity components, but only one spatial position and direction.  
In this class would be included structure functions where
the angle between $\overrightarrow{u}$ and 
$\overrightarrow{\partial_x}$ is not 0 or $90^\circ$ \cite{Aradetal98}.
The different rotationally invariant flatnesses (\ref{eq:Fe4})
involve different velocity and spatial directions, but only one position, $r=0$.
It should be noted that a new type of
structure function that uses two spatial positions in the same
direction, has recently come into use in conjunction with
fusion models \cite{BBiferalefusion}.  These are
similar to the dissipation-dissipation correlation functions
$<\epsilon_x\epsilon_{x+r}>$ where
the two distances would be 0 (the derivative for dissipation)
and the separation $r$ between the two locations of dissipation $\epsilon$. 

While generalized structure functions 
could only be completely determined by numerical simulations
or complicated probes at low Reynolds numbers, some insight
might be gained by considering whether their number could be reduced.
First, a way of systematically
writing these down is needed \cite{Orszag69}.  However,
unlike the way all fourth order derivative correlations
can be expressed in terms of just four irrotational correlations,
no simple reduction to a small number of fourth order
structure functions has been found.  
What can be said is that if this number could be
reduced, it would have to satisfy two rotational groups, one for
position and the other for the velocity components 
\cite{Spector98}.  The full group would correspond to the spin plus angular
momentum group in quantum mechanics.
Even if such a reduction does not exist, 
one would expect some general properties to hold among
all fourth-order structure functions.  For example, if there
is a subrange where the longitudinal and transverse structure functions, 
$S_4^L(r)$ and  $S_4^T(r)$, do have the same scaling,
then maybe all fourth-order structure
functions, including those related to fusion rules, are related in a
similar manner.  Or if there is a subrange where 
$S_4^L(r)$ and  $S_4^T(r)$, do not have the same scaling, then
relationships such as the fusion rules should not apply.

This leads us to three fourth-order structure functions measurable with
crossed wire probes, each corresponding to one of
the three derivative flatnesses (\ref{eq:Fu4}-\ref{eq:Fv4}).
The longitudinal, mixed and transverse fourth-order structure functions are
$$S^L_4=S_{\ell\ell\ell\ell}=<(u(x+r)-u(x))^4>$$ 
\EQ 
S^M_4=S_{\ell\ell tt}=<(u(x+r)-u(x))^2(v(x+r)-v(x))^2>
\label{eq:S4} \EN 
$$S^T_4=S_{tttt}=<(v(x+r)-v(x))^4>$$
whose general moment form is given by eqns. 13.83-13.84 of \cite{MoninY75}.
However, the general moment form in no more fundamental than (\ref{eq:S4})
in terms of the full rotational group.
One would expect that if there
are different scalings for the derivative flatnesses (\ref{eq:Fe4}) under
$R_\lambda$, there should also
be correspondingly different $\zeta^{L,M,T}_p$, with $\zeta^M_p$ 
and $\zeta^T_p$ more strongly dominated by the vorticity statistics.  

\section{Why two length scales? \label{sec:why}}

Another way of looking at the fourth-order velocity structure
functions is to consider the second
order pressure structure function $P_2(r)$.  This is related to a combination
of fourth-order velocity structure functions (\ref{eq:S4}) by
\cite{Hill_Wilczak95} 
\EQ
P_2(r)=<(p(x+r)-p(x))^2>=-{1\over3}S^L_4(r)+
{4\over3}r^2\int_r^\infty y^{-3}[S^L_4(y)+S^T_4(y)-6S^M_4(y)]dy
\label{eq:p2}
\EN
$$+{4\over3}\int_0^r y^{-1}[S^T_4(y)-3S^M_4(y)]dy$$
Assuming $S^L_4(r)\sim\zeta_4^L$,$S^T_4(r)\sim\zeta_4^T$, and
$S^M_4(r)\sim\zeta^M_4$ then
$$P_2(r)=-{1\over3}(1+{4\over{2-\zeta_4^L}})S_4^L(r)
+{4\over3}({1\over\zeta_4^T}- {1\over{2-\zeta_4^T}}) S_4^T(r)$$
$$-4({1\over\zeta^M_4}- {2\over{2-\zeta^M_4}})
S^M_4(r) $$
For the purposes here, the particular numerical prefactors are not
important.  The point to be made is that
if the fourth-order longitudinal and transverse structure functions
have different scalings, then $P_2$ should have different scaling at
the two ends of the inertial subrange.  Assume that
$S^L_4\sim r^{\zeta^L_4}$ and 
$S^T_4\sim r^{\zeta^T_4}$ with 
$\zeta^L_4>\zeta^T_4$ and 
the scaling for $S^M_4$ between these.
Then for $r$ small, $P_2$
would be dominated by the scaling of $S^T_4$ and for large $r$ by
$S^L_4$, with some crossover
length scale in the middle of the inertial
subrange marking the separation between two regimes for the scaling
of $P_2$.  

Related to $P_2(r)$ is the pressure spectrum $S_p(k)$.
It has been suggested \cite{Nelkin99} that the
dependence of $P_2(k)$ on the spectral equivalents
of the dissipation-dissipation correlation
function $<\epsilon_x\epsilon_{x+r}>$ and also $<\Omega_x\Omega_{x+r}>$
and a cross correlation $<\Omega_x\epsilon_{x+r}>$ places a kinematical
constraint upon exponents for these correlations, and therefore
the related fourth-order velocity structure functions.  The relation
used to show this is
\EQ
S_p(k)={{\epsilon^2}\over{4k^4\nu^2}}[E^\epsilon(k)+E^\Omega(k)
-E^M(k)]
\label{eq:P2spec}\EN
where $E^\epsilon(k)$, $E^\Omega(k)$, and $E^M(k)$
are the spectral versions of $<\epsilon_x\epsilon_{x+r}>$, 
$\nu^2<\Omega_x\Omega_{x+r}>$, and $\nu<\Omega_x\epsilon_{x+r}>$
respectively. If each goes as 
\EQ
E^\epsilon(k)=C^\epsilon\epsilon^2k^{-1}(kL)^{-\mu^\epsilon}
\label{eq:Eeps}\EN
with corresponding $\mu^\Omega$ and $\mu^M$ for
$E^\Omega(k)$ and $E^M(k)$, it is then argued that
unless $\mu^\epsilon=\mu^\Omega=\mu^M$ for all $k$ there will be divergences
in $S_p(k)$ as $\nu\rightarrow0$.  This is a reasonable conclusion
for $k\rightarrow0$.  However, for $k>k_\lambda\sim1/\lambda$ where
$k_\lambda^4\nu^2\sim O(1)$, there is no need for this requirement.
Therefore, there could be a kinematical constraint for 
$0<k<k_\lambda$ requiring that $\mu^\epsilon=\mu^\Omega=\mu^x$. 
Relating this to physical space, there would be a constraint
that $\zeta^L=\zeta^T=\zeta^M$ as $r\longrightarrow0$ and
$R_\lambda\longrightarrow\infty$.  This will be shown to be
consistent with the analysis in sections (\ref{sec:newx},\ref{sec:news}). 
For $k_\lambda<k<k_\eta$, 
there would be no constraint, so that it would be possible that
$\mu^\epsilon\neq\mu^\Omega\neq\mu^M$ and 
$\zeta^L\neq\zeta^T\neq\zeta^M$ as $r\longrightarrow\eta$, 
which is also indicated
by the analysis here.  What would be most satisfying would be if
the pressure spectrum itself showed a clear break near $k=1/\lambda$.
This has now been found in the pressure analysis of the $1024^3$ data
to be discussed here \cite{Gotoh00}.

\section{New experimental evidence \label{sec:newx}}

There have been numerous experimental and observation studies of
turbulence designed to determine the scaling of structure functions
and related measures of intermittency such as the dissipation-dissipation
correlation function (\ref{eq:ddiss}).  
However, careful examination raises several questions.
First, while the highest Reynolds number observations \cite{DhruvaTsujiSreeni97}
do find that $\zeta^T(p)\neq\zeta^L(p)$, when ESS is used, compared with
lower Reynolds number results, the difference is noticeably less.  This
suggests that in the very high Reynolds number limit the difference
could go to zero.  On the other hand, ESS might be giving a false impression
of good scaling since the ratios $S^T_p(r)/S^L_p(r)$ as functions of $r$
do not show good scaling behavior to as small a scale as plotting with ESS does.  
A recurrent limitation is that scales below $\lambda$ in
experimental and observational analysis are usually not shown, perhaps
because it is felt that the small scales are less reliable.  For example,
the dissipation-dissipation correlation function exponent is 
expected to be $\mu\approx0.25$ from the observed value of 
$\zeta^L_6\approx1.78$.
This is confirmed by several high Reynolds number measurements 
(\cite{Pras_sp,SreeniKai93} and references therein),
but only for $r\geq\lambda$. It is noted \cite{SreeniKai93} 
that lower $R_\lambda$
experiments tend to give $\mu\approx0.5$.  One moderate Reynolds number
($R_\lambda=400$) experiment for a circular jet \cite{AntoniaSH82} that
shows $\mu\approx0.2$ for $r\geq\lambda$, also shows a breakdown in scaling of 
$<\epsilon_x\epsilon_{x+r}>$ for $r<\lambda$ that could be 
with the origin of $\mu\approx0.5$ at low $R_\lambda$.

\epsfxsize=12cm
\begin{figure}[htbp]
\vspace{10pt}
\epsfbox{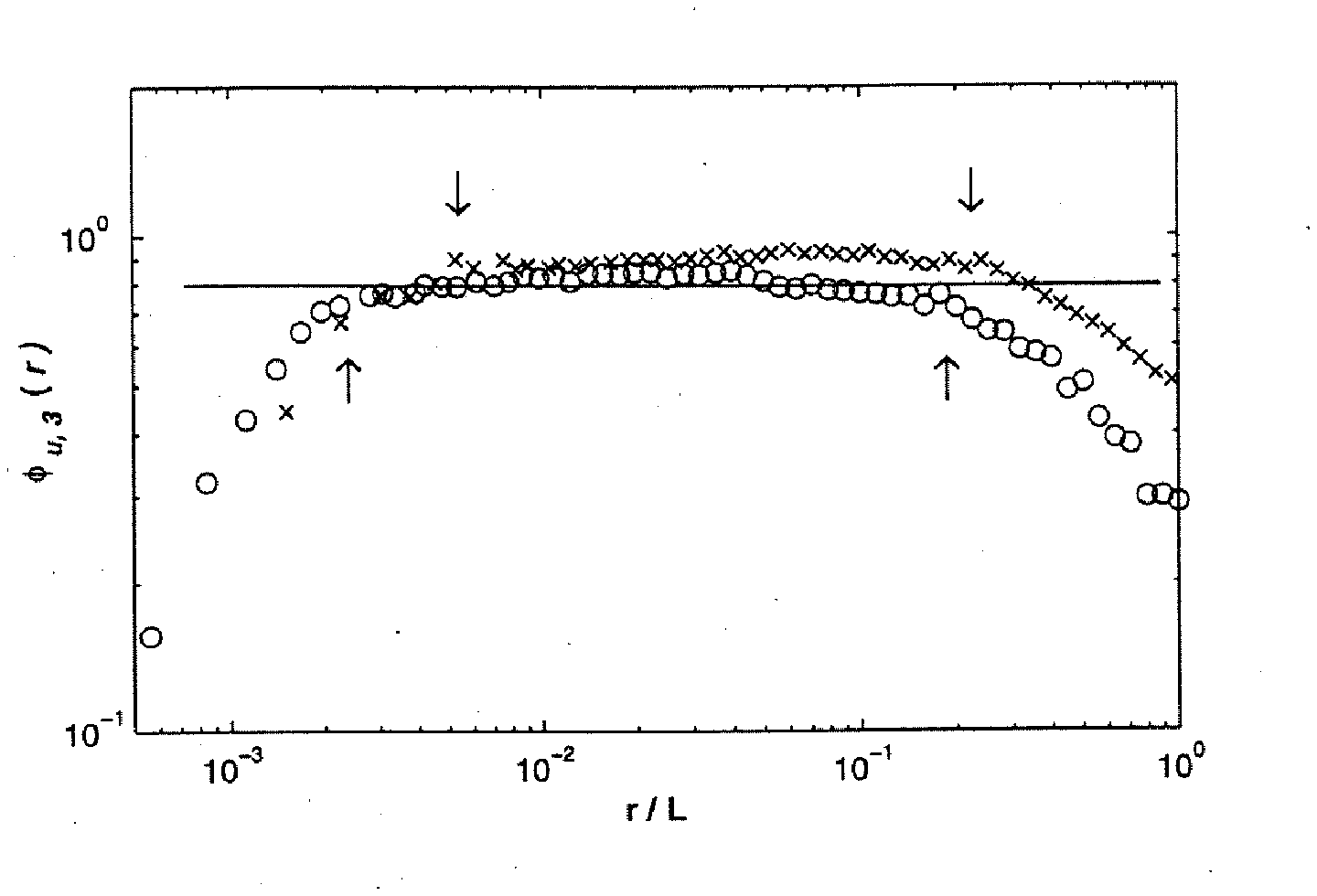}
\vspace{-10pt}
\caption[]{The normalized third-order velocity structure functions.
$\phi_{u,3}=-<\Delta u_r^3>/(<\epsilon>r).$ $\times$, ML; $\circ$, RC.
The vertical down and up arrows indicate the classical inertial-range bounds
($r_\eta$ and $r_L$) for ML and RC,
respectively.  The isotropic value of $\overline{\phi_{u,3}(r)}=0.8$
is shown by the solid line.}
\label{fig:3rd}
\end{figure}
The new experimental analysis comes from data from two experiments
at the Moscow wind tunnel that
has been used to investigate a number of fundamental issues involving
turbulent spectra and structure functions \cite{Pras_sp,PraskovskyPH97}.
The two experiments are 
for a mixing layer ML at $R_\lambda$=2100 and a return channel RC at
$R_\lambda$=3200, which are the highest Reynolds
number laboratory data sets available with both longitudinal and
transverse velocities.  The advantage of an experiment over atmospheric
observations is that an experiment offers more
controlled conditions, which could be especially important for
determining transverse structure functions since any inherent
anisotropy in the flow could affect their values.  With respect to this,
an important point to remember in the following discussion
is that the ML data set is very anisotropic 
and RC data set is nearly completely isotropic.
Therefore, only the RC, $R_\lambda=3200$ experimental data can be
directly compared to the isotropic numerical data to be presented.
The ML data set is included to show that most of the relevant properites
also appear when there is anisotropy.
Extended self-similarity will not be used in order to emphasize
the regimes with simple scaling and where this breaks down
within the inertial subrange.  It will be shown that the scalings
of all of the measured higher-order structure functions change
within the inertial subrange.  

\epsfxsize=16cm
\begin{figure}[htbp]
\hspace{90pt}
\vspace{10pt}
\epsfbox{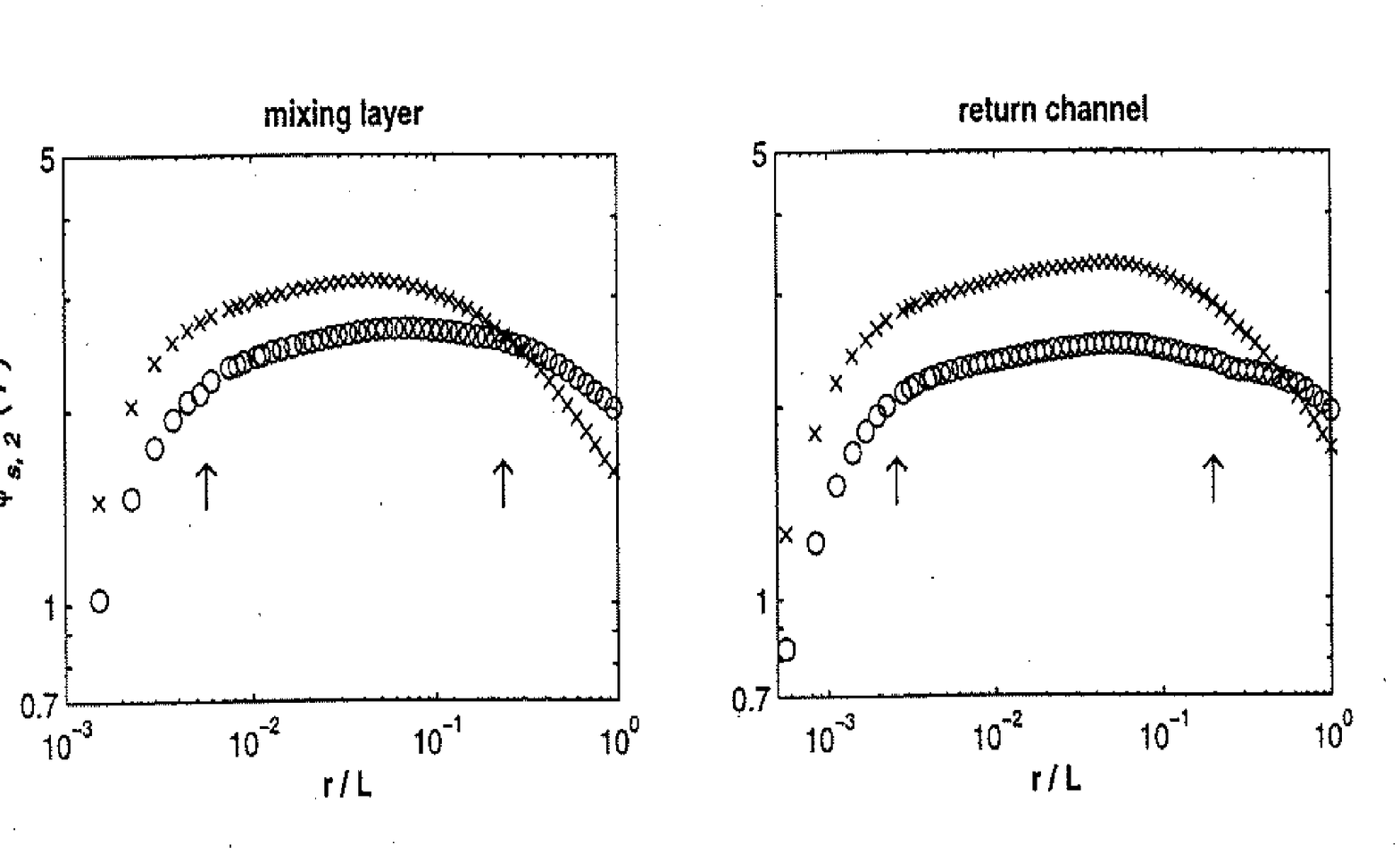}
\vspace{-10pt}
\caption[]{The normalized second-order velocity structure functions.
$\phi^{L,T}_{s,2}=<\Delta s_r^2>/(<\epsilon>r)^{2/3}.$ 
$\circ,~s=u,~\phi_{u,2}=\phi^L_2$; $\times,~s=w,~\phi_{w,2}=\phi^T_2$.
$\eta=0.00018$ and $0.00014$ for ML and RC respectively.  Arrows
are $r_\eta$ and $r_L$.
The vertical arrows indicate the classical inertial-range bounds.}
\label{fig:2nd}
\end{figure}
Figure \ref{fig:3rd} shows the third-order longitudinal structure 
functions $-S_3(r)$ divided by $r$ for ML and RC.  Excluding using the absolute
value, all odd order transverse-only structure functions are zero.
Figure \ref{fig:3rd} uses arrows to indicate 
the inertial subrange based upon the regime over which $-S_3(r)/r$ is constant
for ML and RC.  For ML and RC respectively, at the large scale arrows
are lengths $r_L=L/5$ and $L/6$ and at the
small scales lengths $r_\eta=30\eta$ and $25\eta$.
Figure \ref{fig:2nd} shows the second-order longitudinal $S^L_2(r)$
and transverse $S^T_2(r)$ structure functions for ML and RC.
For $S^L_2(r)$, at large $r$ there is scaling out to an $r_L$
consistent with $r_L$ from $S_3(r)$ in Figure \ref{fig:3rd}.
In addition, there is a new large
length that we will call $r_{4/3}<r_L$ over which the 
isotropic relation $S^T_2(r)\approx (4/3) S^L_2(r)$ holds.
$r_{4/3}$ is smaller than the range over which the longitudinal
second order structure function $S^L_2$ has good scaling behavior,
which is roughly out to $r_L$,
but is consistent with the maximum $r$ for which the transverse
second order structure function $S^T_2$ has good scaling.
For ML and RC, $r_{4/3}=L/30$ and $L/25$  respectively.
The small length scale $r_\eta$ is roughly where the one must
begin to apply extended self-similarity, ESS, (section \ref{sec:constraints})
if one is to get
good scaling relationships for the higher order longitudinal
structure functions down to the Kolmogorov scale $\eta$.
The reason $r_\eta$ is being introduced is to clearly indicate that any new
length scale between $r_\eta$ and $r_{4/3}$ or $r_L$ is fully
within the inertial subrange.
While $r_{4/3}<r_L$ at large scales, the range over which the
4/3 rule for $S^{L,T}_2(r)$ seems to fit at small scales extends to $r<r_\eta$.
This would be consistent with how extended self-similarity
extends scaling regimes more into the dissipation regime.

\epsfxsize=12cm
\begin{figure}[htbp]
\vspace{10pt}
\hspace{90pt} \epsfbox{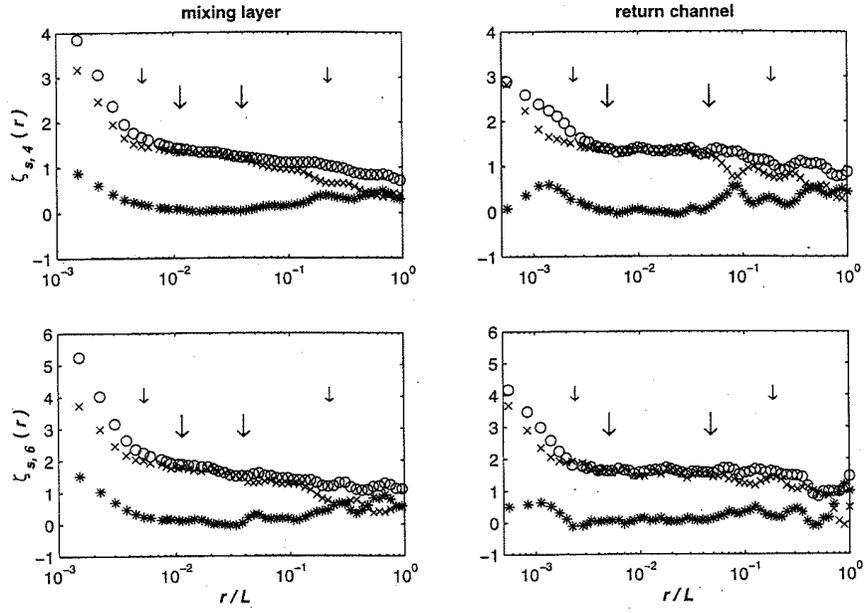}
\vspace{-10pt}
\caption[]{The local slopes $\zeta^{L,T}_p(r)=d\log[S^{L,T}_p(r)]/d\log(r),
\quad p=4,6.$  $s=u$ for $\zeta^L_p$ and $s=w$ for $\zeta^T_p$.
$\circ,~\zeta_{u,p}(r)=\zeta^L_p(r)$; $\times,~\zeta_{w,p}(r)=\zeta^T_p(r)$;
$*,~R_p(r)=S^L_p(r)/S^T_p(r)$.
The outer small vertical arrows indicate the classical inertial
range bounds, $r_\eta$ and $r_L$, while the inner 
large vertical arrows indicated the new bounds, $r_S$ and $r_{4/3}$,
where $\zeta^T_p\approx\zeta^L_p$.  }
\label{fig:4th-6th}
\end{figure}
Figures \ref{fig:4th-6th}(a-d) plot the logarithmic derivatives
of the fourth and sixth-order longitudinal and transverse structure functions.  
For fourth and higher order structure functions there are no
isotropy relationships that would require that $S^T_p$ have
the same scaling as $S^L_p$ over the entire inertial subrange,
only the pressure spectum argument (section \ref{sec:why})
that $\zeta^T_4$ should equal $\zeta^L_4$
for $r>\lambda$.  Consistent with the pressure spectrum argument, 
there is a regime in Figures \ref{fig:4th-6th}(a-d) where the
higher order structure functions do have the same scaling, between
two lengths, $r_{4/3}$ and a new small length scale that we will
call $r_S>r_\eta$.  Referring to Table 1 
for ML ($R_\lambda\approx2100$) and 
RC ($R_\lambda\approx3200$), $r_S=60\eta$ and
$50\eta$ respectively while 
$\lambda=R_\lambda^{1/2} \eta\approx 46\eta$ and $56\eta$ respectively.
So $r_S\approx\lambda$, but it is also roughly twice the value of $r_\eta$.
For $r<r_S$, both $\zeta^T_p$ and $\zeta^L_p$ increase rapidly from their
$r>r_S$ constant values, with $\zeta^L_p$ increasing the
fastest.

This analysis shows a regime within even the rather strict
definition of a measurable inertial subrange defined by
$r_\eta< r< r_{4/3}$ where universal scaling of longitudinal
and transverse structure functions is found and therefore assumptions
of refined self-similarity might apply.  In addition, there is 
a regime for $r<r_S$ where both the longitudinal and
transverse scaling exponents diverge from constant behavior.  Praskovsky's
analysis shows that if scaling functions were fitted over
the entire inertial subrange, that different exponents for the longitudinal
and transverse structure functions would be found that would be
consistent with recent experiments \cite{DhruvaTsujiSreeni97} and simulations
\cite{BoratavP97,Chen_SNC97}.

\section{New numerical evidence \label{sec:news}}

To determine more clearly how significant
the differences between the longitudinal and transverse structure
functions are, it is necessary to see some trends.
For this purpose we now move to analysis of a $512^3$ forced calculation
of isotropic turbulence in a periodic box with $R_\lambda=262$
and another $1024^3$ calculation with $R_\lambda=390$.  The cleaner
numerical data also allows one to more directly compare $S^L_2$ and
$S^T_2$ and to apply extended self-similarity.

The simulations are classic turbulence simulations in a periodic box
with the lowest band of wavenumber modes forced to have constant
energy by Gaussian white noise.  Statistics were taken 
several eddy turnover times after any large excursions 
in the dissipation due to the initial conditions have dissappeared
and once the dissipation rate had settled to the point where it was
varying by less than 5\% on the timescale of several eddy turnover
times.  Experience has
shown that this type of Gaussian white noise forcing yields stable statistics
over a single eddy turnover time.
The $512^3$ simulations
were not dealiased and the statistics represent an average over 40 large
scale eddy turnover times done on the Cray T3D of the
IDRIS Institute in Orsay, France. The
$1024^3$ calculations were done on a Fujitsu VOO5999/56 at the 
Nagoya University Compution Center and represent an average over
1.1 large scale eddy turnover times.  How well the $1024^3$ calculation
satisfies isotropy relations for $S^{L,T}_2(r)$ and $S_3(r)$ 
is discussed elsewhere \cite{Fukayametal00}.
Figure \ref{fig:sp512} is the three-dimensional kinetic energy spectrum
for the $512^3$ calculation, showing a clear
-5/3 regime.  The Kolmogorov scale $\eta$ is half the mesh size 
$\eta=\Delta x/2$.  This is considered
adequate resolution because the peak of the dissipation spectrum, roughly
where the -5/3 regime rolls over into the dissipation regime, is at
$k\approx k_\eta/30$ where $k_\eta=2\pi/\eta$.

In the experiments above, the original definition of an inertial
subrange placed the smallest scale useful for analysis at about 
$r_\eta=25\eta$, roughly the scale associated with the peak of the 
dissipation spectrum.  This definition of
$r_\eta$ would make it greater than $\lambda$ for both simulations,
which for $R_\lambda=262$ is $\lambda\approx16\eta$
and for $R_\lambda=390$ is $\lambda\approx20\eta$.  Therefore,
a quantitative means of choosing $r_\eta$ must be determined that
can be applied equally to simulations at lower Reynolds number and the
$R_\lambda=3200$ experiments.  A difficulty in choosing a method
is that the simulations
are not at high enough a Reynolds number to exhibit either a
clearly defined linear regime in $S_3$ or a long regime where the
isotropy relationship between $S_2^L$ and $S_2^T$ is obeyed.

\begin{figure}[htbp]
\vspace{-20pt}
\epsfbox{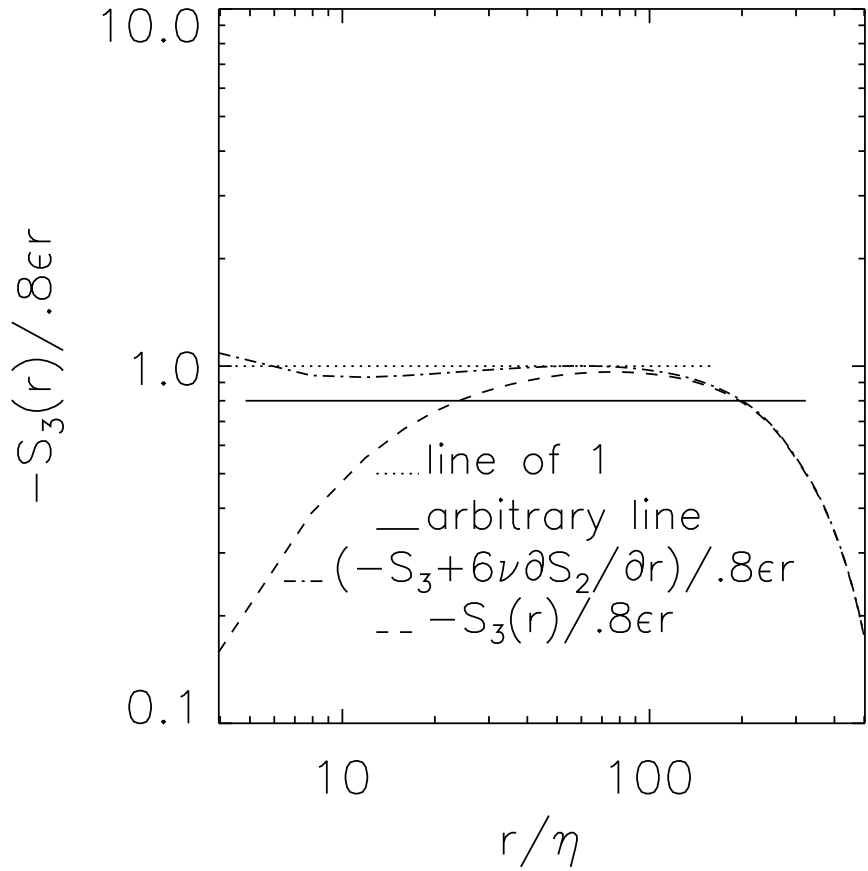}
\caption[]{$-S_3(r)/(.8\epsilon r)$.  
Also shown is
$(-S_3(r)+6\nu\partial S_2^L(r)/\partial r)/(.8\epsilon r)$.
An arbitrary constant line is drawn to indicate where
an inertial subrange is defined. $r$ is with respect to the
size of the box $(2\pi)^2$}
\label{fig:s3512}
\end{figure}

The problem is illustrated by
Figure \ref{fig:s3512}, which plots $-S_3(r)/(.8\epsilon r)$ 
and the normalized form with the viscous correction 
\EQ
D_{LLL}(r)=
\frac{-S_3(r)+6\nu{{\partial S_2^L(r)}\over{\partial r}}}{.8\epsilon r}
\quad.
\label{eq:3rdn} \EN
While the span over which $D_{LLL}(r)$ is 1 indicates that there is an 
energy cascade, an inertial subrange over which
$-S_3(r)/.8r$ is flat as in the experiments in Figure \ref{fig:3rd} 
is not seen.  Therefore, arbrtrarily a inertial subrange will be
chosen as those $r$ where $-S_3(r)/(.8\epsilon r)~>~0.8$.
This is the consistent with how $r_\eta$ was chosen for the RC
experiment and gives consistent values of $r_\eta$ for the
$512^3$ and $1024^3$ simulations and the RC experiment.
Based upon Figure \ref{fig:s3512},
an inertial subrange can be defined between $r/\eta=$19 and 193.

\begin{figure}[htbp]
\vspace{-20pt}
\epsfbox{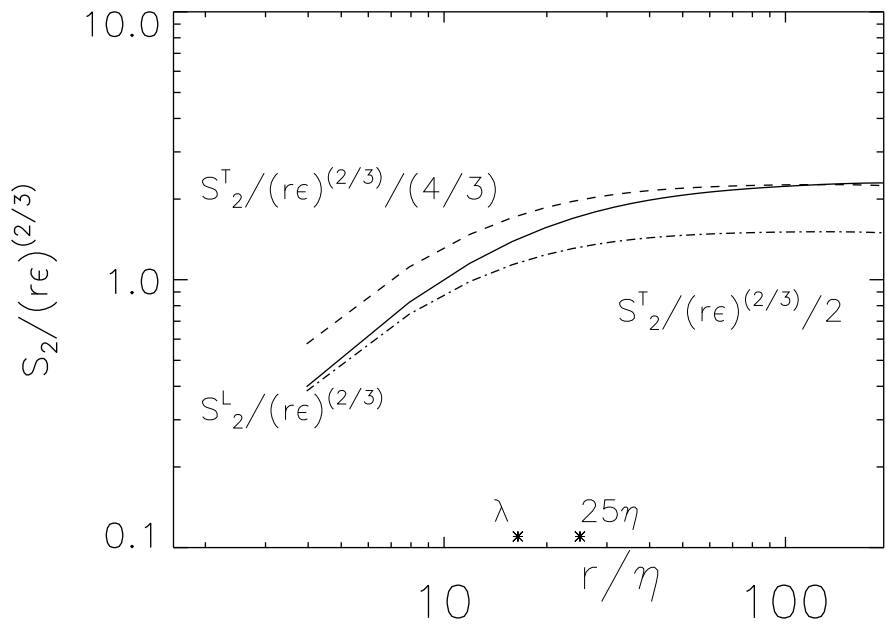}
\caption[]{$S^L_2(r)/r^{2/3}$ (line) and $S^T_2(r)/r^{2/3}$ 
plotted twice.  Once divided by 4/3 to show
the inertial subrange (dash) 
and once divided by 2 to show the dissipation range (dot-dash).}
\label{fig:s2512}
\end{figure}

\epsfxsize=12cm
Let us now use $S^L_2(r)$ and $S^T_2(r)$ to consider this definition
of an inertial subrange using
the isotropic relationship in 3D between the transverse and
longitudinal structure functions.
\EQ
S^T_2(r)=S^L_2(r)+{r\over2}\partial S^L_2(r)/\partial r
\label{eq:s2iso}\EN
In the inertial subrange where $S_2(r)\sim r^{2/3}$, from (\ref{eq:s2iso})
one gets $S^T_2(r)\sim (4/3)S^L_2(r)$.  
As $r\rightarrow0$, $S_2(r)\sim r^2$,
so one gets $S^T_2(r)\sim 2S^L_2(r)$.
Figure \ref{fig:s2512} shows the second-order
longitudinal and transverse structure functions divided by $r^{2/3}$.  
$S^T_2(r)/r^{2/3}$ is plotted twice,
first divided by 4/3 to demonstrate how well the inertial range
relation is obeyed and then divided by 2 to demonstrate the approach
to the dissipation range. 

The 4/3 rule in Figure \ref{fig:s2512} is only approached at the largest
scales, for $r$ greater than where $-S_3(r)/(.8\epsilon r)$ is 
greatest in Figure \ref{fig:s3512}, which could be called $r_{\max}$. 
For $r<r_{\max}$ in Figure \ref{fig:s2512}, the gradient of 
$S^L_2(r)$ is slightly steeper than the $r^{2/3}$ prediction, which
would be consistent with measurements of a small correction over
many years that is usually interpreted as due to intermittency.
However, $S^T_2(r)$ does not show this correction, and since it
is $S^T_2(r)$ which forms the major portion of the energy (4/5ths),
it should not be surprising that the energy spectrum (Figure \ref{fig:sp512})
is very close to -5/3.  If the slope in $S^L_2(r)/r^{2/3}$ is a
dissipation range effect, then one could
compensate for this in analysis of higher order structure functions by
using a new ESS variable based upon $S^L_2(r)$ 
\EQ
R^L_2=(S^L_2(r))^{3/2}/\epsilon,
\label{eq:RL2}
\EN
rather than $-S_3(r)$.  This is done below.
\epsfxsize=12cm

\begin{figure}[htbp]
\vspace{-10pt}
\epsfxsize=12cm
\epsfbox{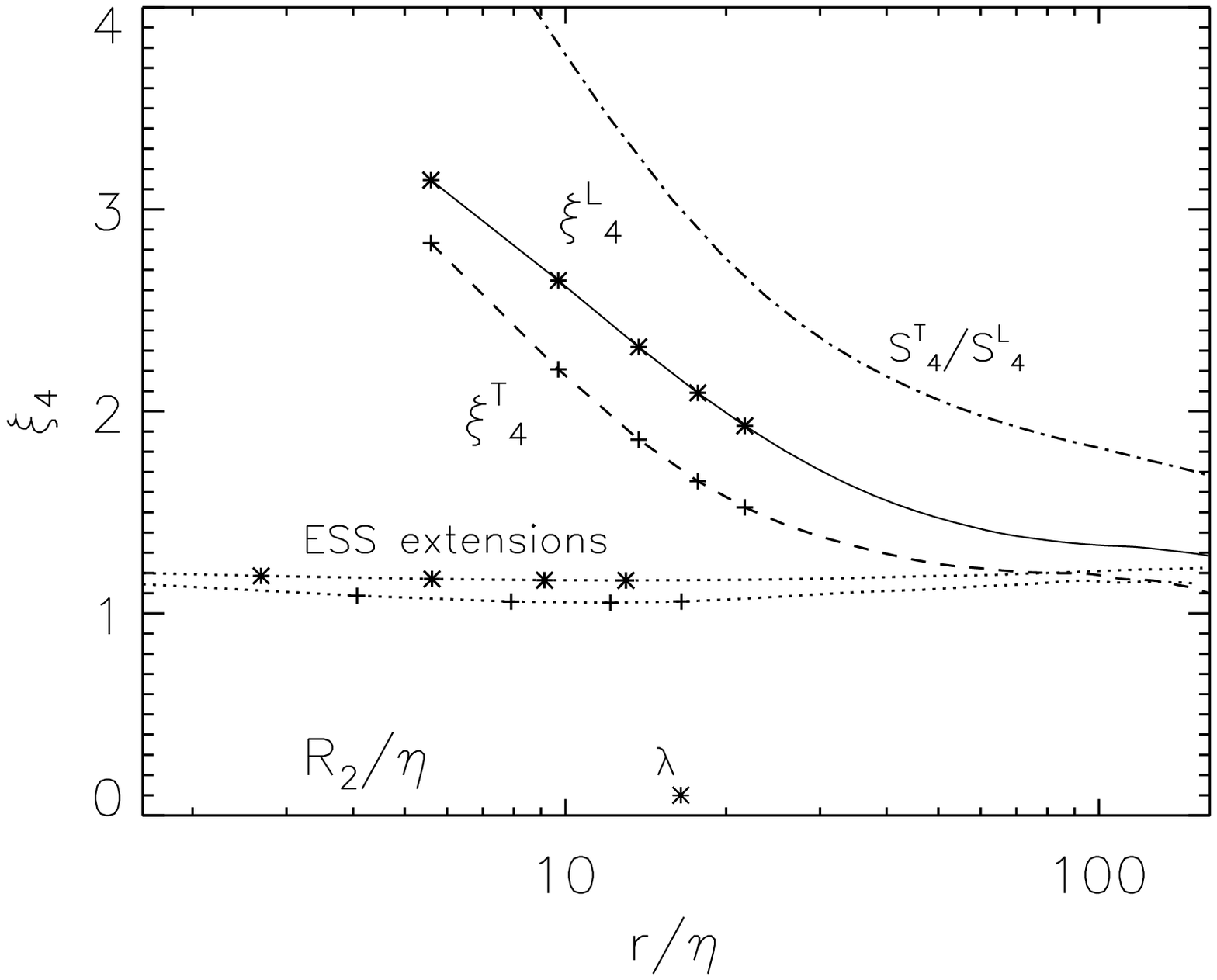}
\epsfxsize=12cm
\epsfbox{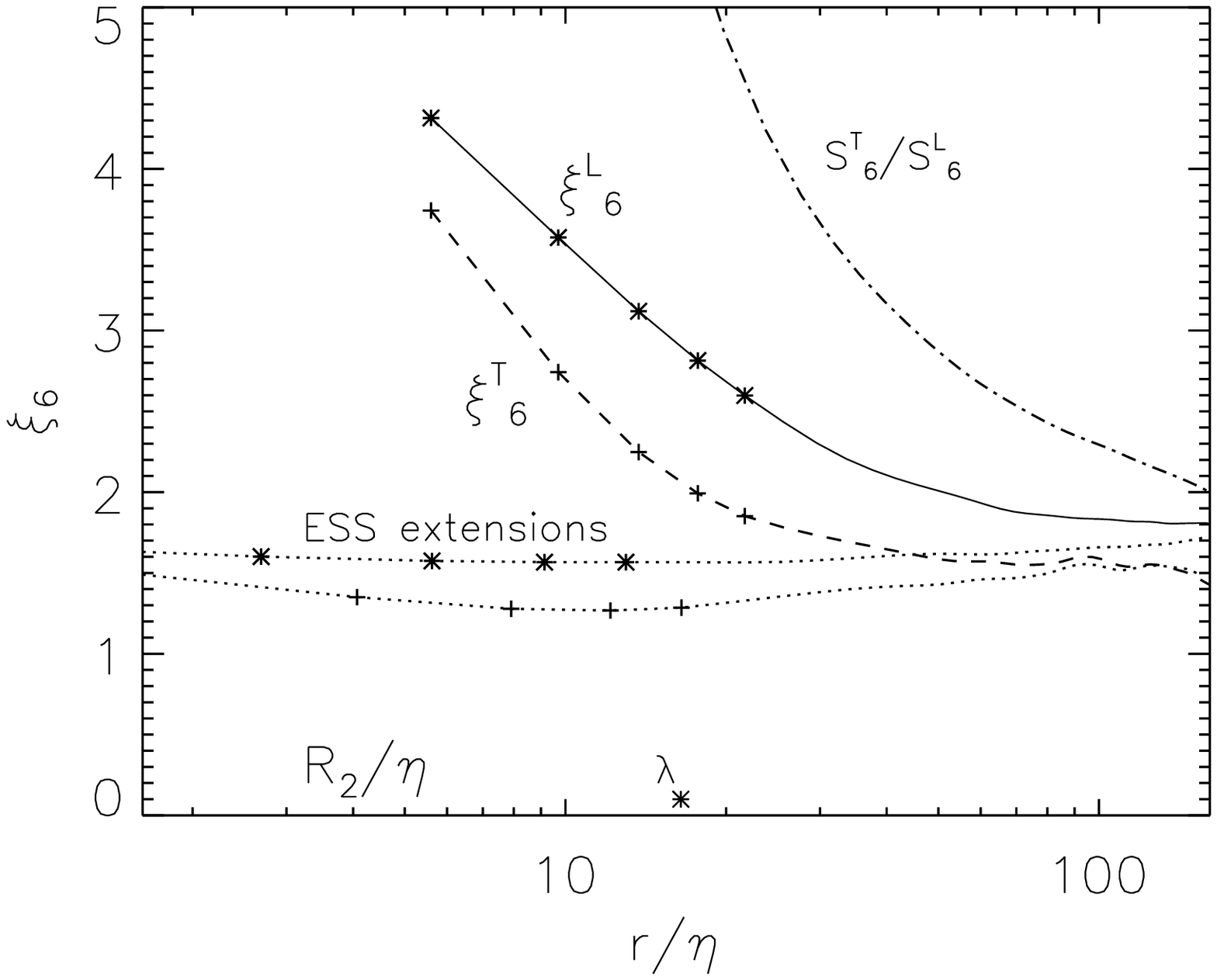}
\caption[]{$\zeta_4(r)$ and $\zeta_6(r)$ for $R_\lambda=262$.
Also shown are $S^T_4(r)/S^L_4(r)$, $S^T_6(r)/S^L_6(r)$ and
extended self-similarity using $R^L_2$ (\ref{eq:RL2}) is applied.}
\label{fig:z4-6512}
\end{figure}

\begin{figure}[htbp]
\epsfxsize=12cm
\epsfbox{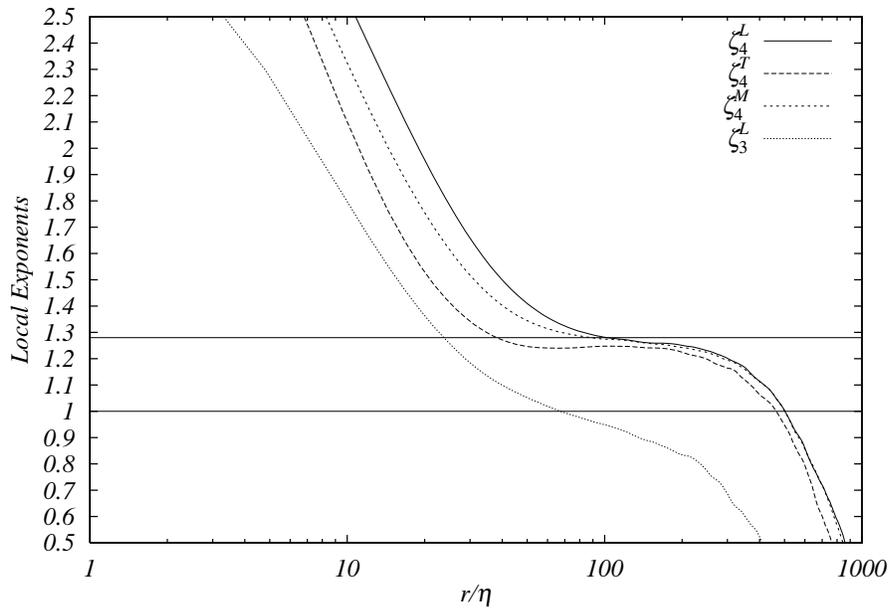}
\caption[]{$\zeta^L_4(r)$, $\zeta^T_4(r)$, $\zeta^M_4$ and
$\zeta_3$ for $R_\lambda=390$.}
\label{fig:z4K3}
\end{figure}

For direct comparison with the experiments, Figures \ref{fig:z4-6512}a
and \ref{fig:z4-6512}b show the logarithmic derivative of the 4th and
6th order structure functions for $R_\lambda=262$ against $r$ and
Figure \ref{fig:z4K3}
shows $S_4^{L,M,T}(r)$ for $R_\lambda=390$.  
As in Figure \ref{fig:4th-6th}, there is a span we will define as 
$r_L>r>r_S$ where the slopes
of the longitudinal and transverse structure functions are constant, but
in this case they are not identical.  In the experiments, $r_S$ was
chosen to be the first $r$ (from below) where $\zeta^T_p(r)\neq\zeta^L_p(r)$.
Since there is no regime where $\zeta^T_p(r)=\zeta^L_p(r)$ in the
simulations, the choice of $r_S$ is more subjective for the simulations.
We have chosen $r_S$ to be where a line of constant $\zeta_4=1.28$ at large $r$
would meet a line with a logarithmic dependence 
through $\zeta^T_4$ at small $r$. The results are given in Table 1
and are similar to the experiments in that $r_S$ is greater than 
$r_\eta$, the lower limit that was defined for inertial range behavior in $S_3$.

What is particularly similar to the experiments in Figure \ref{fig:z4-6512}
is that for $r<r_S$, all $\zeta_p^{L,T}$ increase 
and this increase is greater for $\zeta^L_p$ than for $\zeta^T_p$.  
There is also a clear trend where the difference between $\zeta^T_p$ and
$\zeta^L_p$ defined as
\EQ
\delta_p=\min_r\{\zeta^L_p-\zeta^T_p\}
\label{eq:deltap}
\EN
is decreasing as $R_\lambda$ increases, pointing to the
experiments where $\zeta^T_p\approx\zeta^L_p$ for $r>r_s$.  
\epsfxsize=12cm

\epsfxsize=12cm

\begin{figure}[htbp]
\epsfbox{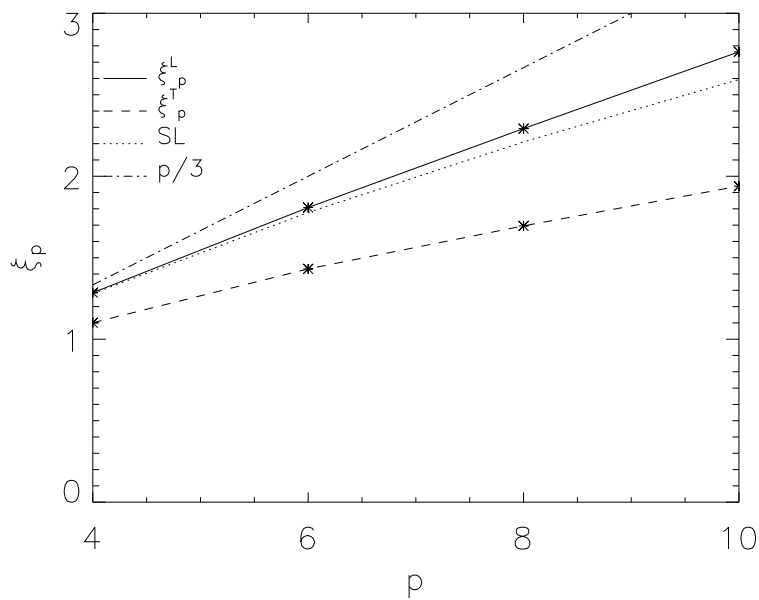}
\caption[]{$\zeta^L_p$ and $\zeta^T_p$ for $R_\lambda=262$.
Values are taken at $r/\eta=188$. SL is the She-Leveque formula
\cite{SheLeveque94}, which is an excellent fit to the observations.
$p/3$ is the classical prediction.}
\label{fig:zetap}
\end{figure}

Applying ESS with $R_2$ brings the slopes back down for $r<r_S$, 
long regimes for scaling $S^L_4(r_3)$ and
$S^L_6(r_3)$ appear, and the differences between $\zeta^T_p$ and $\zeta^L_p$
appear more clearly.  For the $R_\lambda=262$ simulation, 
Figure \ref{fig:zetap} shows different $\zeta^T_p$ and $\zeta^L_p$
that are consistent with earlier work in the same Reynolds number
regime \cite{BoratavP97,vandeWater95,Chen_SNC97}.
However, the long regime of nearly constant
$\zeta^L_p$ should now be considered an artifact of using ESS and
the need to apply ESS seems to be intimately tied to
the new length $r_S$, where $r_S$ is not simply a constant times $eta$.
\epsfxsize=12cm

\section{Discussion \label{sec:conclude}}

In Kerr \cite{Kerr85}, in addition
to velocity statistics, equivalent passive scalar and mixed
velocity-scalar statistics were calculated.  It was noted that
the scaling of the derivative flatness for a passive scalar $\theta$
\EQ
F_{\partial\theta^4}=
F_{\nabla\theta}=(9/5)<(\nabla\theta)^4>/(<(\nabla\theta)^2>)^2
\label{eq:Ftheta}
\EN
was similar to the vorticity flatness $F_\omega$, that is
$z^\theta_4$  \cite{AntoniaChambers80} 
and $z^\omega_4$ were both much larger than the exponent $z^e_4$ for
$F^L_4=F_e$.  There was also a
strong anti-correlation between the scalar derivative and vorticity,
that is
\EQ
F_{\nabla\theta\omega2}=<(\overrightarrow\nabla\theta
\cdot\overrightarrow\omega)^2>/
<(\nabla\theta)^2><\omega^2>~<~1
\label{eq:Fthetaomega}
\EN
Taking this analogy between the statistics of the scalar gradient
and vorticity a step further, if the scaling of $F_{\nabla\theta}$
could be used as a tool for determining the scaling of $F_\omega$,
then high Reynolds experiments for temperature statistics
\cite{Antoniaetal84} could have been implying greater 
scaling exponents in the transverse derivative correlations  and
greater deviations from classical structure function
exponents long before the new work with crossed-wire probes. 
However, without more theoretical understanding and corroborating
evidence from velocity structure functions, 
these hints were not studied further. 

\begin{table}
\begin{center}
\begin{tabular}{lcccccc}
case &  $R_\lambda$ & $\delta_4$ &$\delta_6$ & $r_\eta$ & $r_S$ & $\lambda$ \\
$512^3$ & 262 & .2 & .35 & $27\eta$ & $27\eta$ & $16\eta$ \\
$1024^3$ & 390 & .05 &  & $22\eta$ & $40\eta$ & $20\eta$ \\
ML & 2100 & 0 & 0 & $30\eta$ & $60\eta$ & $46\eta$ \\
RC & 3200 & 0 & 0 & $25\eta$ & $50\eta$ & $56\eta$ \\
\end{tabular}
\caption[]{Dependence of $\delta_p=\zeta^L_p-\zeta^T_p$ (\ref{eq:deltap}),
$r_\eta$ and $r_S$ on case and $R_\lambda$.}
\end{center}
\label{table1}
\end{table}

The theoretical, experimental and numerical discussion here replaces 
that speculation with moderate to high Reynolds numbers experimental 
and numerical data about the behavior of the transverse structure functions.  
The important points are 
that for $p>3$ that $\zeta^T_p\approx\zeta^L_p$ for $r>r_S$
as $R_\lambda\longrightarrow\infty$, where
$r_S$ is further into the inertial subrange than expected, and 
that none of the $\zeta_p$'s show simple scaling
behavior for $r<r_S$. Table 1 
shows $\delta_4$, $\delta_6$, $r_\eta$, $r_S$ and $\lambda$ 
for the numerical cases of
$R_\lambda=262$ and $R_\lambda=390$ and the experimental cases ML
($R_\lambda=2100$) and RC ($R_\lambda=3200$).  Between ML and RC,
mixing layer and return channel, the trend for $r_\eta$ and $r_S$
is opposite that between the much lower Reynolds number simulations
and the experiments.  This is probably just a reflection of the influence
of anisotropy for ML.  The overall trend 
supports the existence of a length scale $r_S$ that is much larger than $\eta$,
with the scaling of $r_S$ as a function of $R_\lambda$ intermediate
between that for $\eta$ and that for $\lambda$.  

The theoretical discussion showed that this second small length scale
would not be expected if the energy cascade and the refined similarity
hypothesis were controlled only by the statistics of the dissipation $\epsilon$.
For there to be a dynamically significant length scale within the inertial
subrange, there must be something
in addition to $\epsilon$ controlling the cascade, the fundamental
dissipation mechanism must involve two length scales, or both.  
This paper addresses only one kind of higher order statistic, the
structure functions.  Clearly a thorough analysis of all higher
order statistics, including pressure and the dissipation-dissipation
correlation function, needs to be done on available measurements and
simulations from the point of view of determining whether some
length scale of the order of the Taylor microscale has a role for
them also.  This has been done for pressure spectra for the $1024^3$
data set used here \cite{Gotoh00} and the results are consistent with
the existence of such a length scale separating spectral regimes of
-7/3 and -5/3.  

Either numerically or observationally, consistent conditions over a wide
range of Reynolds numbers are necessary if any conclusions are to be drawn.
This is difficult to obtain with atmospheric measurements.
As an example, for one series of atmospheric measurements over
a wide range of Reynolds number,
$<\epsilon_x\epsilon_{x+r}>$ has been determined over the entire inertial
subrange \cite{Pras_sp}.  For larger $r$, there is convergence to
$\mu\approx0.2$, but there is an enormous scatter between different measurements
at smaller $r$ so that any $R_\lambda$ dependence in the equivalent
of $r_S$ for $<\epsilon_x\epsilon_{x+r}>$ would be difficult to determine.
Therefore, well-controlled high Reynolds number experiments would be
very useful.  This could also provide a motivation for doing a
careful $2048^3$ forced simulation, which is now feasible.

If confirmed, what could be the dynamical significance 
of this crossover length scale 
beyond just being an average between the integral scale
and the Kolmogorov scale?  To date, no importance has been attached
to the Taylor microscale $\lambda$ beyond that of an average.  When the first
visualizations of vortex filaments were done \cite{Kerr85}, 
one way of characterizing them was they had a width the order of the
Kolmogorov microscale and a length the order of the Taylor microscale.
However, in a $64^3$ or $128^3$ DNS, it would be impossible to determine
whether the length was $\lambda$, a fraction of the size of
the box, or just a multiple of $\eta$.  That is,
the the radius of curvature might just be a multiple of $\eta$.  
The highest resolution visualizations of
isotropic, homogeneous turbulence \cite{PorterPW94} would support this
scenario.  That is, vortex filaments are observed, but they do not
snake through the entire domain and instead have a length that appears
to be a only a multiple of $\eta$.  However, it can be argued that
these are hyperviscous calculations that are predisposed to shortening 
the vortex filaments.  Furthermore, statistical models based strictly
upon vortex filaments \cite{SheLeveque94,HeLosAlamos98}
do not seem capable of producing different longitudinal and transverse scaling
in the high Reynolds number limit. 

Therefore, it seems that
some other type of dynamical object besides simple filaments
would be needed if a theoretical basis for a second dynamically significant
length scale is to be given.  The only dynamical structure that has been
identified in either turbulence or idealized calculations of Navier-Stokes
and Euler that self-generates two small length scales is the structure found
in the interaction of two anti-parallel vortex tubes as the peak vorticity
appears to be developing a singularity \cite{Kerr93}.  However, the spectrum of
this structure is $k^{-3}$, nowhere near $k^{-5/3}$. So until high Reynolds
number, very highly resolved calculations are done for Navier-Stoke vortex
reconnection, any connection between the properties of this structure and 
the scaling properties of turbulence is pure conjecture.

Acknowledgements.  NCAR is supported by the U.S. National Science
Foundation.  R.M.K. wishes to thank Service d'Astrophysique, Centre
d'Etudes de Saclay for support.  M.M. wishs to thank the Centre
National de Recherche Scientifique of France for computing support.
T.G. thanks deeply to Nagoya University Computation Center for 
supporting the computation and to the support of the  
Grant-in-Aid for Scientific Research (C-2 09640260) 
by The Ministry of Education, Science, Sports and Culture of Japan.

\end{document}